\documentstyle[11pt,aaspp,psfig]{article}

\begin{document}

\title{Two-Dimensional Hydrodynamics of Pre-Core Collapse: Oxygen Shell Burning}

\author{Grant Baz\'an \altaffilmark{1}}
\author{David Arnett\altaffilmark{2}}
\affil{Steward Observatory, University of Arizona,
   Tucson, AZ 85721}

\altaffiltext{1}{e-mail: bazan@dirac.as.arizona.edu} 
\altaffiltext{2}{e-mail:  darnett@as.arizona.edu}


\begin{abstract}
By direct hydrodynamic simulation, using the Piecewise Parabolic Method (PPM)
code PROMETHEUS, we study the properties of a convective oxygen burning shell
in a SN 1987A progenitor star ( 20 M$_{\odot}$) prior to collapse.  
The convection is too heterogeneous and dynamic to be well approximated
by one-dimensional diffusion-like algorithms which have previously been
used for this epoch. Qualitatively new phenomena are seen.

The simulations are two-dimensional, with good resolution in radius
and angle, and used a
large (90-degree) slice centered at the equator. 
The microphysics and the initial model were carefully
treated. Many of the
qualitative features of previous multi-dimensional simulations
of  convection
are seen, including large kinetic and acoustic energy fluxes, which are not
accounted for by mixing length theory.  Small but significant amounts of  
$^{12}$C are mixed non-uniformly into 
the oxygen burning convection zone, resulting in hot spots of 
nuclear energy production which are 
more than an order of magnitude more energetic than the 
oxygen flame itself.  
Density perturbations (up to 8 \%)  occur at 
the ``edges" of the convective zone and are the result of gravity waves 
generated by interaction of penetrating flows into the stable region. 
Perturbations of temperature and Y$_e$ (or neutron excess $\eta$) 
at the base of the convective 
zone are of sufficient magnitude to create angular inhomogeneities  
in explosive nucleosynthesis products, and need to be included in
quantitative estimates of yields.  Combined with the plume-like 
velocity structure arising from convection, the perturbations will
contribute to the mixing of $^{56}$Ni throughout supernovae 
envelopes.  Runs of different resolution, and angular extent, 
were performed to test the robustness of these simulations.
\end{abstract}


%
%

\section{INTRODUCTION}

Oxygen burning in a convective shell is 
one of the last, dominant burning stages before 
core collapse of a massive star.  Its importance lies not just in the 
nucleosynthesis occurring by the nuclear burning itself, but 
also in the structure 
left by convection which might affect subsequent supernova evolution 
(\cite{a69a}, \cite{fa73},
\cite{afm89}).  For instance, the division between the proto-neutron star and  
supernova ejecta, known as the ``mass cut,"  is expected to appear 
somewhere near the ``edge'' of the convective zone.  
Convection generated perturbations in density, 
velocity, entropy, and Y$_e$ (or neutron excess $\eta$) could 
alter the nature of core collapse as compared to spherically symmetric, 
one-dimensional calculations (\cite{a77a}, \cite{bl85}, 
\cite{br85}, \cite{mw91}; see \cite{a96} for extensive discussion and references).

The bottom of the oxygen convective shell is expected to be the region
where explosive nucleosynthesis occurs.  Most iron peak isotopes are thought to
be produced here.  The most important individual product of this region would be
$^{56}$Ni, which is responsible for powering supernova light curves  
and (after decay to Co) is the
source of the 847 and 1208 keV $\gamma$-ray lines seen in  SN1987A 
(\cite{ls90}, \cite{t90}, \cite{k92}, \cite{p93}, \cite{aq92}).  The explosive 
O- and Si-burning which produces $^{56}$Ni also  
produces other $\gamma$-ray emitting radionuclides  such as $^{44}$Ti and  
$^{57}$Ni ($^{57}$Co).  Perturbations in Y$_e$, density, and temperature, left by convective 
O-burning, will {\em greatly} affect the relative abundances of individual 
isotopes, due to the sensitivity of nuclear statistical  
equilibrium (NSE) and quasiequilibrium
 to these parameters (\cite{tcg66}, \cite{wac73}, \cite{a96}).   
Studies of NSE have shown 
a neutron excess $\eta = 1 - 2Y_e \approx 0.002$ best recreates 
the relative abundances of iron-peak nuclei, with slight changes in $\eta$ 
altering relative abundances by large factors.  A 
change in $Y_e$ from 0.499 to 0.4995 changes the relative abundance of 
$^{57}$Fe to $^{56}$Fe by roughly a factor of 2.

We know from  observations of SN 1987A that significant mixing of newly
synthesized material thoughout the envelope must occur 
(\cite{a88}, \cite{pw88}, \cite{k89},  \cite{g88}) .  The only physical
explanation that agrees with our understanding of massive star evolution is 
that significant density perturbations must appear at the interfaces of  
composition discontinuities and these result in Rayleigh-Taylor
(and Richtmeyer-Meshkov) mixing in the 
ejecta (\cite{fa73}, \cite{bt90}).  This hypothesis has been tested with various 
fluid hydrodynamics codes by adopting ad hoc distributions of
initial perturbation amplitude 
and length scale (\cite{afm89}, \cite{fma91}, \cite{mfa91}, \cite{a91},
\cite{hmns92}, \cite{hb92}, \cite{hw94}).  On the whole, 
calculated $^{56}$Ni velocities
were in agreement with observations for the bulk of the matter, but not for
the highest speeds. Not enough Ni could be mixed to velocities as high 
as 3,000 km $\rm s^{-1}$, which the observations imply.  We believe that 
convective shell O-burning has a major effect on all of these phenomena.

In almost all hydrostatic stellar models, convection is treated by
enforcing an adiabatic gradient, or better,
some version of mixing length theory (\cite{p25}, \cite{v53}, \cite{bv58}; 
hereafter MLT).  The basic tenets of MLT are that (1) convection occurring deep
in stellar interiors results 
in an almost completely adiabatic  structure, and that (2) excess energy flux 
(above that of radiative diffusion at the adiabatic gradient) is a function 
of the local 
superadiabatic gradient.  The mixing of chemical species is approximated either 
by complete homogenization,  or the use of a radial  diffusion  equation 
(when MLT formalism is applied).   The 
concept of penetrative convection (overshoot from convective regions  into 
adjacent radiative  zones) is not a natural outcome of MLT, due to the local 
nature of the theory (\cite{r87}).  Various ad hoc approaches to overshoot 
have been applied to stellar evolution; in most the overshoot distance is 
scaled to some multiple of the local pressure scale height.

Multi-dimensions simulations of turbulent compressible convection have revealed
both the successes and shortcomings of MLT. One dynamical feature of convection,
which has been present in every multi-dimensional simulation, is the asymmetry
between up and down motions.  Upward moving elements are typically broader in extent and
slower than downward moving elements, although the degree of asymmetry is
very dependent on the density contrast of the unstable region (\cite{g75},
\cite{htm84}, \cite{csw82},  \cite{htm86}, \cite{cs86}, \cite{cbtm91}).  As 
density contrast increases, so does the asymmetry 
(\cite{htm84}).  Pressure fluctuations  (the compressible 
nature of the fluid) can assist in driving descending motions, lead to  
buoyancy braking in ascending motions,
and satisfy a Bernoulli relation with horizontal velocities 
(\cite{htm86}).  The basic notion of buoyancy driven motions 
throughout most of the convective zone has been upheld, but at the 
boundaries, dissipation and compressibilty are the dominant 
factors in the flow.  In the more complex case, in which convective
regions have nuclear timescales comparable to the mixing timescale, 
a kinetic (rather than steady state) description becomes crucial.

There are hints that the basic tenet in MLT regarding {\em energy} flux
still holds true in multidimensional simulations.  In both 2- and 3-D ``large
eddy" simulations, in which a sub-grid scale turbulence formalism (SGS) 
has been applied, 
vertical motions have been seen to breakup in less than the simulation dimension 
(\cite{csw82}, \cite{cs86}, \cite{cs89}, \cite{hm91}).  In fact, the 
autocorrelation of vertical velocity along vertical direction is a function of 
the difference in logarithmic pressure, but symmetric with a similar 
full width-half maximum (FWHM) 
across the entire computational regime (\cite{csw82}, \cite{cs86}).  This 
means that the convective flux should be a function of the local pressure 
scale height.  Amazingly, vertical velocities seem to be correlated over about 
one pressure scale height (\cite{hm91}), which is the range used for the past 
30 years in stellar evolution models.   However, even these positive
indications are faulty. 
There seems to be a 
numerical dependence to these results:  the correlation is no longer 
constant over an entire grid when the aspect ratio is increased (\cite{cs86}) 
and the correlation length is dependent on the magnitude of the SGS turbulence 
included in the model.  Indeed, in models where SGS turbulent viscosity has 
been omitted in favor of ordinary dynamical viscosity, no breakup of 
vertical motions has been observed (\cite{htm84}, \cite{htm86}).

Penetrative convection has also been modelled in multi-dimensions.  In what 
follows, we will use a nomenclature in which {\em penetration} 
refers to mixing (in formally stable zones)
that is efficient enough to weaken thermal stratification.  Otherwise, 
edge mixing is 
referred to as {\em overshooting} (\cite{mltz84}).  Fast downward plumes are 
responsible for extensive mixing of composition and energy from the unstable 
region into the lower stable region through gravity waves (\cite{htm86}), 
although a rapid increase in the stability of adjacent regions reduces the 
ability of plumes to penetrate (\cite{htmz94}).  Composition is advected  
across the stability boundary to a ``penetration depth,''
on the timescale for plume transit. In the case of true penetration, the region 
adjacent to the convective
zone can be described by two layers: (1) where Peclet number $> 1$, adiabatic 
structure exists, and motions decelerate due to buoyancy breaking and (2) 
where Peclet number $< 1$, temperature
 perturbations damp out from radiative damping 
(\cite{z91}, \cite{htmz94}). 
Note that the Peclet number is the ratio of the velocity scale $v$ times the
length scale $\ell$, to the thermal conductivity $K$, 
$Pec = v \ell/K$, and a large Peclet number implies
ineffective thermal energy transport relative to mass transport; see \cite{ll59}.
Our case, in which cooling is by 
$\nu\overline{\nu}$ emission instead of radiative diffusion, does not seem to 
have been considered by the hydrodynamic community.

Shell oxygen burning differs from the canonical problem
addressed by most previous multidimensional convection simulations.  
The convective regions of ZAMS A and F stars (\cite{ltm81}, 
\cite{sc84}) and generic convection with ionization (\cite{rt93,rt93a,rt93b})
are most comparable.  
However, such convective regions do not contain nuclear 
energy sources and neutrino energy sinks within the 
flow.  It seems that local thermal effects, 
such as heating from ionization, can significantly affect convective flows.  
The treatment of convection 
affects neutrino cooled regions, since the efficiency of 
convection determines which adiabat a convective region can achieve.  This in 
turn feeds back upon the neutrino emission (\cite{auf93}, \cite{a96}).

In previous work (\cite{a94}, \cite{ba94}), we used our Piecewise 
Parabolic Method fluid dynamics code PROMETHEUS (\cite{afm89}) to 
begin an examination of the problem of the evolution of a  O-burning
shell in a 20 M$_{\odot}$, Z = 0.007 star, prior to core collapse. 
Significant density perturbations (of the order of 7\%) were found to occur at 
the boundaries of the convective region, along with non-uniform mixing of 
chemical species within the shell.  Here, we present more extensive
results of models of 
the same star, with different boundary conditions, grid sizes, and grid 
resolutions, and for a longer evolutionary time.  We compare results for 
different boundary conditions to  see the effects on the dynamics (\cite{hm93}).
Computational domains are  varied in order to see the effects of sound waves on
the system.  Calculations  with different grid resolutions are performed to 
estimate the role of scale lengths on the problem. A detailed analysis of the
simulations and their astronomical implications is given.

\section{COMPUTATIONAL DETAILS}

\subsection{The PROMETHEUS code}

The hydrodynamics code PROMETHEUS is based upon the piecewise-parabolic
method (PPM, \cite{cw84}).  The method constructs the physics of the flow 
between grid points by a non-linear solution of the equations 
for conservation of mass, 
momentum, and energy (the Riemann problem) rather than the usual Taylor 
expansion about the grid points.  This procedure affords greater resolution per 
grid point, which is highly desirable for multi-dimensional problems in stellar 
astrophysics.  Although the computational effort required per grid point is 
higher than other commonly employed numerical methods, the number of grid 
points needed for a given level of accuracy is less (often much less).  The net 
result for PPM is a {\it lower} total computational effort for given accuracies 
than competing numerical methods.  Because the computational load per grid 
point is greater, a host of physics packages (nuclear reactions, radiation, 
gravity, etc.) may be added before affecting the runtime significantly.  Thus 
PPM is well suited for multi-dimensional problems involving significant physics 
beyond the bare hydrodynamics.

The PROMETHEUS code was originally written by Bruce Fryxell and Ewald 
M\"uller, and is an extension of the basic PPM scheme in that (1) it 
includes an arbitrary number of separate fluids to account for abundances of 
nuclear species, (2) nuclear reactions and energy generation are taken into 
account, and (3) a realistic (not a gamma law) equation of state is used 
(\cite{cg85}).  Variations of this original code have been widely 
employed by other 
groups (e.g. \cite{bl93}, \cite{bf92}), including one specifically for 
the study of convection in three dimensions (\cite{cbtm91}).  PROMETHEUS can 
support different grid geometries, with cartesian, cylindrical, or spherical 
coordinates in 1, 2, or 3 dimensions as options.  There is also an option of a 
moving grid, which allows a `semi-lagrangian' approach to certain problems and 
adds greater resolution for problems where flows are significant across the 
entire grid, as in explosions.

\subsection{Equation of State}

The equation of state is the sum of components for electrons, ions, 
and radiation.  The electron contribution is determined via cubic 
interpolation in tables for the thermal contributions 
(including $e^+e^-$ pairs), plus degeneracy.  
Derivatives of the electron pressure and energy 
as functions of temperature and density are obtained from the direct 
differentiation of the interpolation functions.  Table entries are 
logarithmically spaced both in temperature (in degrees Kelvin,
$\Delta$log(T) = 0.05 from 7 $\leq$ 
log(T) $\leq$ 10) and in density (in $\rm g cm^{-3}$, 
$\Delta$log($\rho$) = 0.125 from 2 $\leq$ 
log($\rho$) $\leq$ 10.5).  A perfect gas equation of state is 
smoothly joined for ($\rho$,T) regions not covered by the electron EOS table, 
and used for the ion EOS (to which a coulomb correction is added: \cite{ll69}). 
Radiation pressure and energy density are included.

For the temperatures encountered in quiescent oxygen-burning, nuclear 
statistical equilibrium effects may be ignored, and the effects of 
quasi-equilibrium seem small.  These effects are both ignored with regard 
to the equation of state.

\subsection{Nuclear Reactions}

The algorithms for nuclear reations are similar to those used in 
prior quasi-static stellar evolution studies of massive stars (\cite{a71}).  
Twelve species (e$^-$, H, n, $^4$He, $^{12}$C, $^{16}$O, $^{20}$Ne, $^{24}$Mg, 
SiCa, 
$^{56}$Ni, $^{56}$Co, $^{56}$Fe) are considered for the energy generation and 
nucleosynthesis resulting from pp, CNO, helium (3$\alpha$), carbon, neon, 
oxygen, and silicon burning stages.  The decays of $^{56}$Ni and $^{56}$Co are 
also included with density dependent decay rates (\cite{ffn85}).  The effects of 
{\em steady state} electron captures during oxygen and silicon burning are 
included; Urca processes and their attendant dynamic effects will be dealt with
in subsequent work.  
The approximations in the reaction network used to represent oxygen and silicon 
burning are based upon extensive reaction network simulations (\cite{ta85}, 
\cite{at85}, \cite{a96}).

Cooling by neutrino-antineutrino pair emission is 
included.  This process has been identified in 1D stellar models 
(\cite{a71}) as competing with nuclear burning in a fashion that determines the 
thermal balance in the oxygen convective region.  The thermal conditions of the 
flame zone, in particular, were determined by the bulk of the material in the 
convective region, which is $\nu\overline{\nu}$ cooled, and advected into the flame 
zone.  Using MLT, the heated material would then rise and cool by expansion and
further $\nu\overline{\nu}$ cooling.  Thus, cooling by neutrino-antineutrino pair emission 
is never comparable locally to nuclear heating (\cite{a69a}, \cite{a71}), rather 
it is the integral effect of heating and cooling across the convective zone
which is important for thermal balance.  
We employ the pair emission rates of \cite{bps67}, with small revisions
for neutral current effects (\cite{it96}). 

\subsection{Computational Differences}

There are differences between these simulations and
our exploratory work
(\cite{a94}).  Here we map the structure of the star
resulting from one-dimensional hydrostatic calculations
 onto the upper and lower boundaries of the grid.  Our previous work
utilized constant entropy and isothermal outflow as the respective lower and
upper boundary conditions.  We calculate the 2D hydrodynamics from the
outer region of the Si core to the H envelope, instead of just the O shell
layer.

Another difference is the calculation
of models with periodic boundary conditions.  The initial work, as well as
some of the simulations discussed here, were calculated with reflective boundary
conditions, which are known to attract entrained flows  and enhance the
downward kinetic energy flux (\cite{cs86}). Aside from not being realistic
for sectors smaller than $0 \le \theta \le \pi$,
reflective boundary conditions can also damp waves of wavelengths larger than 
the dimension of the calculation.  Because we do not know
 exactly how power cascades to 
smaller scales in astrophysical conditions, as compared  to box simulations 
with lower Reynolds' number, we find it to be prudent to use periodic
boundary conditions rather than reflecting ones.

\begin{table*}
\begin{center}
\begin{tabular}{rrrrrrr}
Model &  n$_r$ & n$_{\theta}$ & $\Delta r$ & $\Delta \theta$
& B. C.\tablenotemark{1} \\
\tableline
   A  & 256  &  32 & 6.4e+10  & $ \pi \over 2$  & P\\
   B  & 256  &  64 & 6.4e+10  & $ \pi \over 2$  & P\\
   C  & 256  &  64 & 6.4e+10  & $ \pi \over 2$  & R\\
   D  & 460  & 128 & 6.4e+10  & $ \pi \over 2$  & P\\
   E  & 256  &  96 & 6.4e+10  & $3\pi \over 4$  & P\\
\end{tabular}
\end{center}
\caption{Some Models in this Study.}
\label{tbl.1}
\tablenotetext{1}{Boundary conditions: R and P denote reflective and periodic boundaries.}
\end{table*}

We have calculated models over a range of initial grid dimensions, grid 
resolutions,
and boundary conditions to see how changes in any of these quantities would
change the most critical results of this study.  Table~1 shows the various
descriptive qualities of the models. The differences are small (see Appendix).
Most of the discussion is illustrated by the high resolution run, 
case D (460x128). A dominant constraint in choice of zoning is that
the oxygen flame zone be well resolved (\cite{a94}).

\subsubsection{Numerical Effects}

As a check of the robustness of our calculation, we have also run cases
which have the same resolution, but a larger or smaller physical dimension. 
Calculations
with a smaller physical domain are more sensitive to boundary conditions
(and errors), so that larger domains are preferable. This is especially true
for the oxygen shell burning epoch, because the other parts of the star are
also evolving. The Si core is treated as an inert lower boundary; numerical
experiments (to be reported separately) indicate that this approximation is
valid over at least the early part of oxygen shell burning. The outer boundary
is placed near the outer edge of the He core ($M_r \approx 6 M_\odot$).
This insures that the acoustic flux is not artificially trapped, includes
any back reaction from overlying burning shells, and allows us to construct
a reasonable two dimensional approximation to a SN1987A progenitor model.

Cases A, B, and D of Table~1 provide a basis in which to examine effects 
of grid resolution; even the coarse case A gave qualitatively similar
results. Case E tested the effect of a larger angular wedge (135 degrees
instead of 90); the results were similar to case B.

The side boundaries provide a more serious challenge. Reflecting boundaries
tend to trap waves, so that we have used periodic boundary conditions. 
Case C is identical to case B except for the use of reflecting instead of
periodic boundaries; the effect is small for our simulations.

As an attempt to understand possible geometric peculiarities caused by the 
choice of grid geometry, we have used both 
an $(r,\theta)$ coordinate system (Table~1), and some 
cases which used an $(r,\phi)$ coordinate system (\cite{ba94}).
The qualitative nature of the results is unaffected.

\section{DISCUSSION}

Convection in oxygen shells of massive stars is not driven
in quite the same manner as convection in stellar envelopes or terrestrial 
phenomena. The equations to be solved look much like those
of other compressible convection simulations,
\begin{eqnarray}
\frac{\partial\rho}{\partial t}           & +  \nabla \cdot (\rho {\bf v}) & = \ \ \  0, 
\\ \cr
\frac{\partial(\rho {\bf v})}{\partial t} & +  \nabla \cdot (\rho {\bf v}{\bf v}) 
    & =      - \nabla p + \rho {\bf g} , 
\\ \cr
\frac{\partial (\rho E)}{\partial t} & + \nabla \cdot ( \rho {\bf v} E ) 
    & =  - \nabla \cdot ( p {\bf v} ) - K(T,\rho) \nabla T \cr
& & \ \ \   +  \epsilon_{nuc}(X _i,\rho,T) - \epsilon_\nu (\rho,T)
\end{eqnarray}
where $E$ is given by,
\begin{equation}
E = \varepsilon + \frac{1}{2} v^2,
\end{equation}
and $ K(T,\rho)$ is the radiative conductivity.
Notice, however, that there are source and sink terms for nuclear 
energy release and neutrino losses.  Thus, we must solve a set of continuity
equations for composition as well,
\begin{equation}
\frac{\partial(\rho X_i)}{\partial t} + \nabla \cdot (\rho X_i {\bf v}) = 
\rho \left(\frac{\partial X_i}{\partial t}\right)_{nuc} ,
\end{equation}
where, except for the factor of $\rho$, 
the right side is a system of reaction network equations.  
This means that convection in oxygen shells involves energy production and loss
which is directly in the flow.

To our knowledge, the most physically similar system to  
have been examined previously with multi-dimensional simulations is
the core helium flash (\cite{cd80}, \cite{dw87}, \cite{d96}).
The oxygen shell differs in that (1) the reaction rates involved have 
even more extreme temperature dependencies (roughly $ T^{40}$), (2) the 
equation of state is dominated by radiation and electron thermal pressures
rather than degeneracy and thermal pressures, and (3) 
convection is driven by the competition between heating by nuclear burning
and cooling by expansion and by neutrino emission
(rather than cooling by radiative diffusion, as 
is the normal assumption in convective systems).

\subsection{Timescales}

The timescales of the problem portray this connection between convective mixing and
energetics.  Figure 1 shows the various timescales estimated from the MLT solution 
from the initial one-dimensional, hydrostatic stellar evolution model.  
The upper plot 
shows that the convective turnover time across a given radial grid zone,
\begin{equation}
t_{cnv} = \frac{\Delta x_i}{v_{cnv}} ,
\end{equation}
is very much shorter than either the nuclear energy timescale,
\begin{equation}
t_{nuc} = \frac{ E_{th} }{|(\epsilon_{nuc}(X _i,\rho,T) - \epsilon_\nu (\rho,T))|} ,
\end{equation}
where $E_{th}$ is the thermal component of the specific energy, 
or the Kelvin-Helmholtz timescale,
\begin{equation}
t_{KH} = \frac{\Omega}{L},
\end{equation}
where $\Omega$ if the gravitational potential energy and $L$ the luminosity
(radiative and convective). This means that local convective response is fast
enough to control heating and cooling.
However, using MLT velocities,
 the integrated convective turnover time across the whole oxygen 
convective region  is long (10$^4$ seconds, or 550 sound travel times), 
so that the time to move material across
the whole region is approaching both the nuclear timescale at the
base of the convective shell (10$^{4-5}$ seconds), and the neutrino 
Kelvin-Helmholtz timescale (10$^{5-6}$ seconds) through
the oxygen shell.  This implies heterogeneity and significant nonspherical
variation. The timescale for energy flow by radiative diffusion,
\begin{equation}
t_{rad} = \frac{\rho ~ E_{th}}{|\nabla  \cdot (K(T,\rho) \nabla T)|}
\end{equation}
is orders of magnitude larger than all of these timescales, and, thus, is not
important in this stage of evolution.  For this reason, we need not calculate
the effects of radiative diffusion in these simulations.  

\begin{figure}[!hb]
\psfig{file=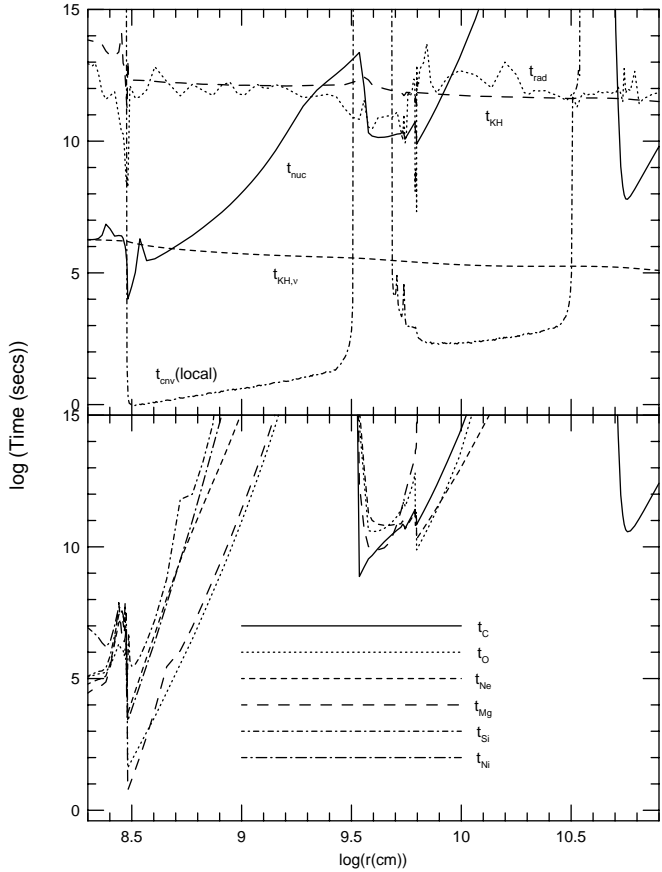,height=5.5in}
\label{fig.1}
\caption{Timescales for Convection and Burning. Mixing and burn times 
are comparable.}
\end{figure}

The lower plot shows
the various abundance timescales, for nucleus $i$, where
\begin{equation}
t_{i} = X_i {\bigg / \bigg \vert} \frac{\partial X_i}{\partial t}{\bigg \vert}.
\end{equation}
The burning timescale of oxygen
 is of the same order as the fastest convective timescale across a
given grid zone, but orders of magnitude less than the turnover time
across the convective region.  Notice that many other nuclei have short 
timescales as well.

The magnitudes of these timescales lead us to several conclusions. 
First, {\em composition will
not be homogeneous.}  This will feed back into the structure via the energy 
source terms.  Second, {\em the system has strong fluctuations in space
and in time.}
Because the thermal energy timescale is of the order of both
the Kelvin-Helmholtz timescale and full convective mixing timescale, the
dynamics of convection and evolution of the star will be 
strongly varying.  This hardly satisfies the requirements of MLT which mandate
that the dynamics of convection reach a statistical steady state.  The energy
supply changes as fast as convection can move energy across the region.

\subsection{Evolution of the Velocity Field}

Even at the beginning of our simulations,
we see that MLT does not correctly represent the dynamics of 
convection in oxygen shells.  Because the fluid transport timescales are 
initially 
longer than the oxygen burning time in the flame zone, the oxygen flame 
undergoes a initial thermal pulse, much like an AGB star (but faster), 
and a sound 
wave is emitted.  This sound wave takes 18 seconds to cross the oxygen shell
and effectively erases the MLT velocity field which we imposed initially.  
This allows the shell to
establish dynamical structures from very low noise levels and reduces any 
worry that the initial perturbations somehow determine the final dynamical 
state. This pulse is not due to a faulty mapping of the initial model onto
the 2D grid, but to an inconsistency between the assumed MLT physics of the 
convective shell, and the actual 2D hydrodynamics with heating and cooling.
It is due to a thermal inconsistency involving assumed and actual convective
cooling, not an error in overall hydrostatic balance.

\begin{figure}[!hb]
\label{fig.2}
\psfig{file=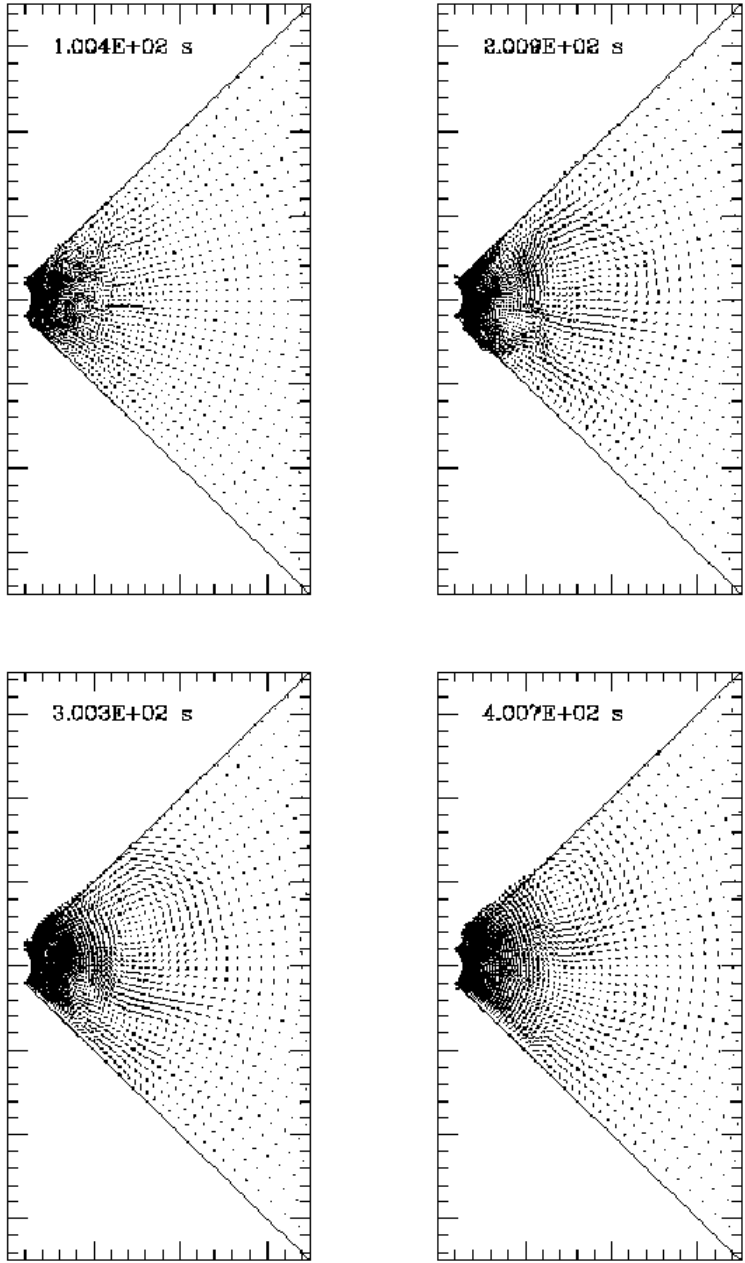,width=5.5in}
\caption{Snapshots of Velocity Fields, 100-400s. Large eddies dominate.}
\end{figure}

Figure~2 shows snapshots of velocity evolution after times of 100, 200, 300, and
400 seconds.  
Before about 100 seconds, the dynamics are dominated by rising and
descending plumes, which are driven by simple buoyancy.  Small eddies (scale 
length $\leq$ pressure scale height) form at the base of the shell and 
form/merge like those in most two dimensional convection simulations (\cite{pw94}).
By 200 seconds (11 sound travel times across the original oxygen convective zone), 
the region is fully convective and buoyancy braking is readily
apparent as a retarding force to upward motions, causing swirls. 
A large eddy dominates the upper two-thirds of the convective region, and one eddy
dominates in the lower third.  At 300 seconds, a single downward
plume spans the upper 2/3 of the zone, and buoyancy still dominates the flow 
dynamics.  By 400 seconds, buoyancy braking affects the dynamics not only at 
the boundaries, but also throughout the zone. 

These simulations exhibit many of the generic traits of multidimensional 
compressible turbulent convection simulations.  
A comparison of these vortices with other two-dimensional convection 
simulations (\cite{htm86}, \cite{pw94}) shows that they are qualitatively 
similar in dimension (length/width $ \sim 1$), and that a few large eddies
dominate.  \cite{pw94} showed that this is to be expected,
as two-dimensional simulations exhibit the merging of counterrotating eddies
into larger ones.  Earlier in our simulations, the lower $1/3$ of the 
domain was populated by small eddies.  

Velocity magnitudes in these eddies are 
typically at Mach numbers of 0.1 to 0.2, which is more than a factor of ten
 higher than MLT would predict.  The spatial 
extent of these eddies is on the order of 3 pressure scale heights.
The simulation at this epoch is still time dependent so that
its structures should not be compared to the mixing
length used in MLT.  Only the radial velocity correlation length of a 
statistical steady state (\cite{csw82}, \cite{cs89}, \cite{pw94}) should be 
used as some measure of a mixing length.  As we explain later, the dynamical
coupling of the energy generation to the kinematical mixing, as well as the
limited number of convective turnover times involved in the entire stage of
shell oxygen burning, prevents a statistical steady state from ever being 
attained in this stellar evolution stage.  

In light of these results, {\em we view the details of
one dimensional stellar evolution calculations of 
the structure of the oxygen shell with skepticism.}
The smooth, steady flow assumed in those simulations
appears to be a phenomenon which does not survive in more realistic
calculations.

\subsection{Energy Fluxes}

The convective energy fluxes in this stage show both similarities and 
differences in comparison to previous compressible convection simulations.  
Figure 3 shows the 
azimuthally averaged enthalpy, kinetic, acoustic, and radiative fluxes 
after 400 seconds. 
The top panel shows the inner part of the O convective shell; the bottom
panel gives the outer part, on different scales. 
Notice that the fluxes in the outer region are
smaller in amplitude by more than a factor of 20.
 The enthalpy flux pictured here is {\sl negative}
through much of the Schwarzschild unstable region; MLT assumes it is positive.
The formal convective boundaries are indicated as vertical dashed lines.
The convective fluxes extend beyond these limits.

\begin{figure}[!hb]
\label{fig.3}
\psfig{file=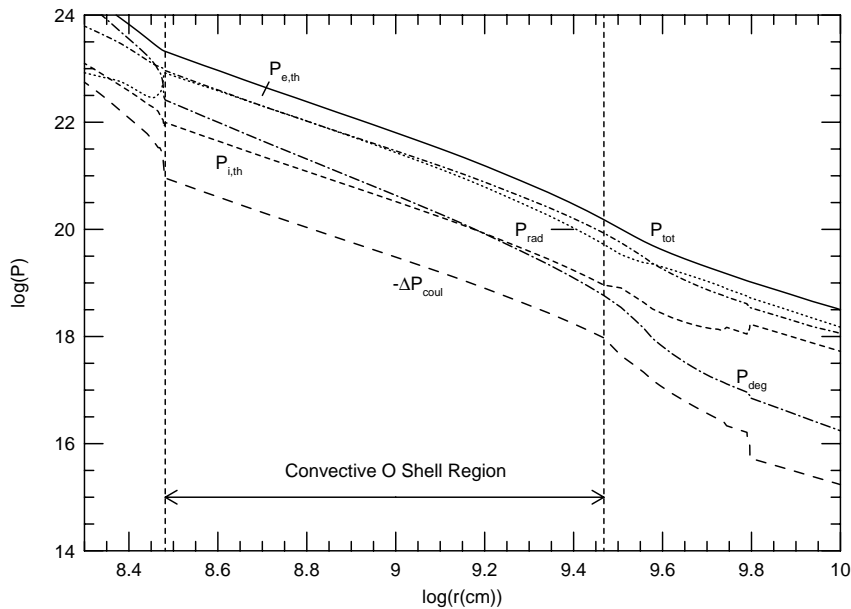,height=3.5in}
\caption{Pressure Contributions vs Radius. Thermal electron and radiation
pressure dominate.}
\end{figure}

As in previous convection simulations,
we find a non-zero flux of energy to be carried in the form of sound waves,
\begin{equation}
F_p = \overline{v_r ~\cdot ~\delta P} ,
\end{equation}
and kinetic energy,
\begin{equation}
F_k = \frac{1}{2}\bigg (\overline{\rho \bf{v ~\cdot ~v} v_r} \bigg ).
\end{equation}
MLT conventions require that the net radial kinetic energy flux be zero and 
that pressure equilibrium is always in effect. However, 
compressible convection simulations (\cite{htm84}, \cite{csw82}, \cite{cs86}, 
\cite{cbtm91}, \cite{pw94}), reveal a finite kinetic energy flux due to cool, 
dense material being compressed into plumes.  Since these plumes occupy a 
filling fraction at constant radius which conserves mass flux relative to 
slower, upward moving material, the kinetic energy flux must be nonzero.  In 
this particular model, the kinetic energy flux can reach as much as 20 \% of
the enthalpy flux, which is less than the peak values of 60 \% seen in 
previous ``box'' simulations.  
One explanation for this difference might be that simulations
of convection
in a box are governed by the competition between convective instability 
and viscous dissipation by radiative (or thermal) diffusion.  
Here, convective motion competes 
with neutrino losses, as well as numerical dissipation. 
The estimated radiative flux is at least
 4 orders of magnitude less than
the enthalpy flux at all points in the simulation.  This fact was 
emphasized in early stellar evolution studies of massive stars (\cite{a72}), 
where MLT theory yielded the result that the luminosity 
at the boundary of such convective zones would be zero. 

\begin{figure}[!hb]
\label{fig.4}
\psfig{file=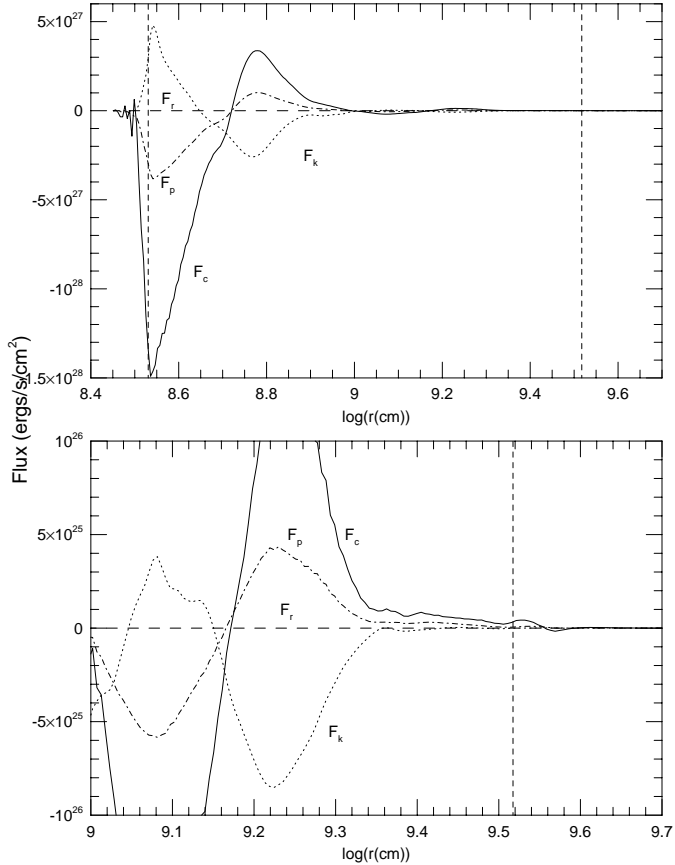,height=5.5in}
\caption{Energy Fluxes at 400s. Acoustic (p), kinetic energy (k), 
enthalpy (c), and radiative (r) fluxes are shown. Enthalpy flux
is often {\em negative}, and acoustic and kinetic fluxes are important.}
\end{figure}

Figure 4 shows the enthalpy and
kinetic fluxes and their sum at the top, middle, and bottom of the convective 
shell as a function of time.  The dot-dashed curve is the kinetic flux, the
dashed curve the convective flux, and the solid curve their sum.
At no time during the simulation is the total luminosity equal to 
zero in the formally stable region.  We note in passing that 
it is fortunate that neutrino losses
govern the extent of this convection, since the numerical dissipation of the
PPM method at this resolution (\cite{pw94}), while yielding effective Reynolds' 
numbers ranging from 10$^8$ at the base of the shell to 10$^5$ at the top, also
imply Prandtl numbers ranging from 200 at the base to 1000 at the top.
The small numerical dissipation is still larger than the tiny radiative dissipation.
This suggests that we would need at least a tenfold increase in resolution before the
radiative diffusion could be quantitatively evaluated, if at this
tiny level it were still important. 

An important component of the energy flow is the
acoustic flux.  This quantity measures the energy carried in the form of
sound waves and is relevant to the kinetic energy flux in
 flows of significant Mach number (\cite{htm84}).  
In places the acoustic flux is a significant
fraction of the kinetic flux.  Thus, {\em any attempt to model systems
such as these with implicit or anelastic methods will miss the 
significant contribution to energy transport by sound waves}.

The enthalpy flux distribution is negative
through much of the Schwarzschild unstable domain.  This is because
the neutrino losses are dominant globally, causing contraction.  
After the first sound crossing time, the simulation adjusts to a state 
representative of the physics involved.
The negative enthalpy flux implies that the energy arising from the oxygen 
burning flame zone is insufficient to suppport the structure predicted by 
MLT, and energy is being drawn from above. Neutrino-cooled matter is contracting
down on the flame zone.

We find that there is a time dependance in many quantities 
throughout the duration of the simulation.  For example, significant deviations
in flux about 
an average value are seen in Figure~4.  However, there may be a trend toward
to a roughly constant amplitude for each of the three regions after 200s.  

There are several reasons for the time variability in the simulation, and for 
it to continue until the time of 
core collapse.   First of these are the similarities of both the nuclear and 
convective turnover timescales to the time left before core collapse.  We 
estimate that there should be no more than 12 turnovers from the birth of 
the convective oxygen shell until core collapse, as estimated from our 
one-dimensional model.  Previous box models of convection certainly required 
more turnovers before reaching a time-independent state (\cite{htm84}, 
\cite{htm86}).  The other reason for time dependence in the simulation has to 
do with the density contrast and the aspect ratio.  Previous work stressed that 
for density contrasts of 21, a minimum aspect ratio (horizontal /vertical 
dimensions) of about 3 was necessary for a static state to be reached in 80 
turnover timescales (\cite{htm84}).  In fact, since our simulation covers 
such a large spherical geometric area, the aspect ratio ranges from $\sim 0.2$ 
at the base to $\pi / 2$ at the upper edge of the convective zone.  Even for a 
$2\pi$ simulation in $r$ and $\phi$, rather than $r$ and $\theta$, the ratio 
is only four times this value.  Simulations still should be performed to see
if these trends still hold in arbitrary geometries (cylindrical + spherical) as
the effects of adiabatic cooling on individual mass elements is more 
enhanced in a curvilinear coordinate system than a cartesian system.  
Nevertheless, assuming the trends from the cartesian simulations roughly hold 
true in a spherical system, and that a density contrast of about 200 exists 
across the convective zone, and the values of our aspect ratios,
our simulations suggest that {\em we can hardly 
expect this convective oxygen burning shell to ever reach a dynamical steady 
state before core collapse} and that {\em the application of mixing length 
theory in convective shell oxygen burning is unjustified.}

\subsubsection{Penetration and Overshoot}

We now consider the presence of 
overshoot in these calculations.  The structure below and above the unstable 
region is not changed much from the initial state, hence the designation of 
{\em overshoot} and not {\em penetration}.  
As Figure 3 showed, thermal energy is 
transported out of the stable region while kinetic energy is transported 
into the region, apparently not in quantities sufficient to alter the structure 
in these regions.  However, the structure of the temperature and density at the
lower boundary is significantly altered.  Most of this has to do with the
mixing timescale being slower than the local oxygen burning timescale,
in which case the oxygen is depleted and Y$_e$ drops locally due to electron
capture on the oxygen burning products.  This, in turn
causes changes in the entropy and composition, which then feeds back into the 
temperature and density.

\begin{figure}[!hb]
\label{fig.5}
\psfig{file=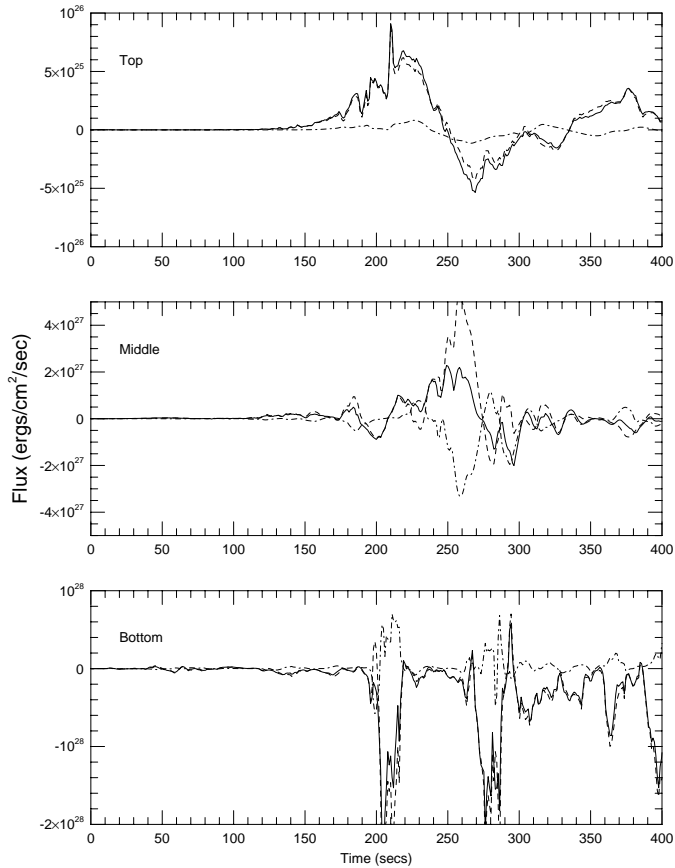,height=5.5in}
\caption{Energy Fluxes: time variability. Even with this low noise initial 
state, convection is well established after 200s, but highly variable.}
\end{figure}

{\em While these mass motions (Reynolds stresses) are not sufficient to alter the 
stellar structure significantly, they do allow mixing of composition from 
the convectively stable region into the unstable region.} Figure 5 shows the 
composition profiles of $^{12}$C.  The $^{12}$C 
abundance in this oxygen-burning convective shell is purely the result of 
undershoot from the upper, $^{12}$C-rich regions. 
A blob of $\rm ^{12}C$ rich matter sinks down through the
oxygen shell, finally halted by its own rapid burning. This is a new effect, not
seen in one-dimensional simulations.
We will discuss the consequences below.

Unlike the box simulations (\cite{htm86}, \cite{htmz94}), 
we find that overshoot is more vigorous at the upper rather than  the lower 
edge of the unstable region.  This seems to be related to the 
relative stability of our upper and lower zones (\cite{htmz94}).  
The presence of a nuclear shell at the bottom of the zone leads to a value 
of $\gamma$ which changes
faster as a function of position than it does at the top of the shell.
This means that the $relative$ stability of the bottom of the zone is higher
than the top of the convective zone. 

\subsubsection{Perturbations}

The early appearance of $\gamma$-lines from decay of $^{56}$Ni,
as well as its effects on the light curve of SN
1987A, left theorists with the idea that
mixing occurred, so that density and/or velocity
perturbations must exist in the progenitor models (\cite{cl87}, 
\cite{a88}, \cite{k88},
\cite{pw88}).  As hydrodynamical simulations have shown (\cite{afm89}, 
\cite{fma91}, \cite{mfa91}, \cite{hb92}, \cite{hbc92}, \cite{hmns92}), these 
perturbations facilitate the mixing of $^{56}$Ni toward the surface through 
Rayleigh-Taylor instabilities.  The exact details of the perturbations, such
as scale distribution and amplitude, have not been self consistently examined,
but linear analysis indicates that instabilities are likely at 
composition interfaces (\cite{bt90},  \cite{mfa91}).  

Because of the behavior
of the equation of state, as seen in its linearized form:
\begin{eqnarray}
\frac{\delta P}{P} &=& \frac{\delta \rho}{\rho} \bigg ( \frac{P_{i,th} + P_{e,th} + 
                     \frac{4}{3} P_{deg}}{P} \bigg ) \cr
\cr
&&+ \frac{\delta T}{T}  \bigg ( \frac{P_{i,th}
                     + P_{e,th} + 4 P_{rad}}{P} \bigg )  \cr
\cr
&&+ \frac{\delta Y_e}{Y_e}  \bigg (
                     \frac{P_{e,th} + \frac{4}{3} P_{deg}}{P}  \bigg ) \cr
\cr
&& + 
                     \frac{P_{i,th}}{P} \bigg ( \sum_j \frac{Y_j}{\sum_k Y_k} 
                     \frac{\delta Y_j}{Y_j} \bigg ) ,
\end{eqnarray}
the nonspherical perturbations are correlated, and
show a strong temperature dependence due to the radiation pressure.

\begin{figure}[!hb]
\label{fig.6}
\psfig{file=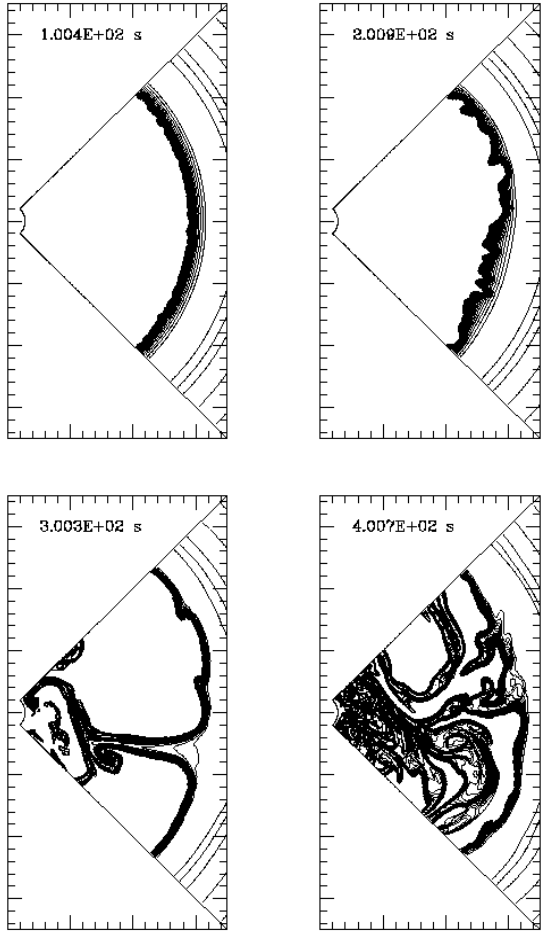,height=5.5in}
\caption{$^{12}$C Plume Snapshots. Such effects are not in 1D calculations.}
\end{figure}

Figure~6 shows the different contributions to the pressure, as a function of radius.
Electron degeneracy and thermal ion pressure contribute
little to the total.  Roughly 60 \% is from 
thermal electrons, while almost all of the remaining 40 \% is radiation pressure.
This simplifies the above expression to one of the form:
\begin{equation}
\frac{\delta P}{P} = \frac{\delta \rho}{\rho} \bigg (\frac{P_{e,th}}{P}  \bigg )+ 
                \frac{\delta T}{T}\bigg ( \frac{P_{e,th} + 4 P_{rad}}{P} \bigg )+ 
                     \frac{\delta Y_e}{Y_e}\bigg ( \frac{P_{e,th}}{P}  \bigg ) .
\end{equation}
This expression is a good description of the fluctuations in the flows.

\begin{figure}[!hb]
\label{fig.7}
\psfig{file=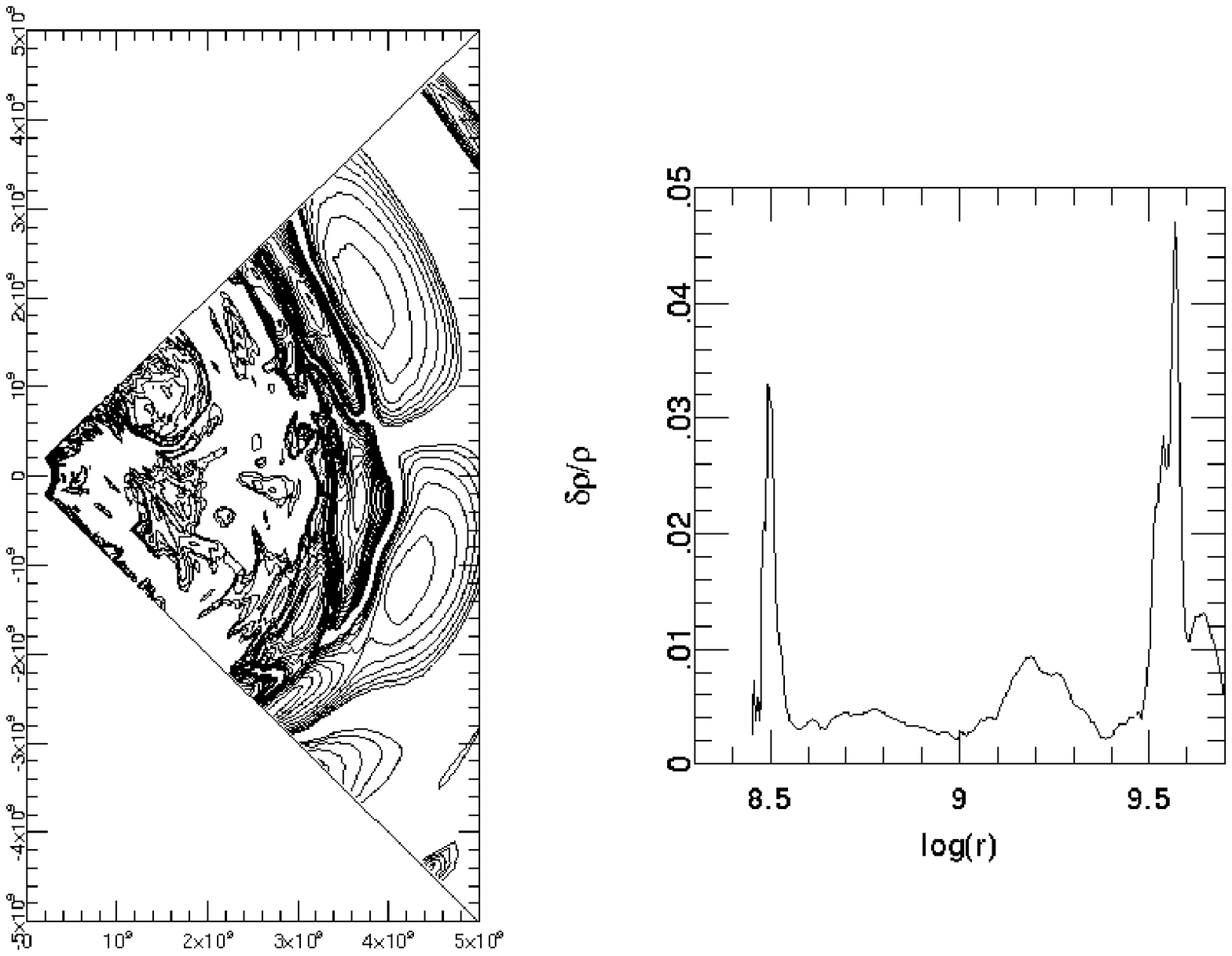,height=5.5in}
\caption{Density Perturbations at 400s. Left: Contours to $\pm10\%$. 
Right: azimuthal averages (rms).}
\end{figure}

In Figure~7, we show the density 
perturbation structure in and around the oxygen shell at the end of the 
calculation. The perturbation contours range over $\pm 10\%$.
The largest perturbations are in the flame zone (the inner convective
boundary), and at the outer convective boundary, as may be seen in the
right-hand panel. Notice that the thin flame zone, which lies near the
leftmost arc in the left panel, has large perturbations in its small
physical volume. The perturbations at the top of the convective zone
have a much larger length scale, although their angular scales are
similar to those in the flame zone. 
The ``perturbation amplitudes'' implied by the 
convection in the oxygen shell easily exceed the minimum needed to give significant 
mixing in supernova ejecta models (i.e. a few percent).  Not surprisingly, the 
highest amplitude perturbations are found at the point where the linear 
analysis indicated: at the composition interfaces.  

This may be sufficient to explain the mixing of most of the  $^{56}$Ni 
observed in SN1987A, but apparently {\em not} that at highest velocity 
($v \approx $ 3,000 $\rm km \ s^{-1}$; see \cite{a91}, \cite{hb92}). 
The production of this high velocity material may be related to the passage 
of the explosion shock, and explosive oxygen burning
in this inhomogeneous shell (\cite{a94}, Bazan and Arnett, in
preparation). It might also be due to mixing outward by overshoot of $\rm^{56}Ni$ 
just prior to explosion (\cite{hb92}).  This would require very high temperature
oxygen burning; the silicon burning core has too large a neutron excess to produce
$\rm^{56}Ni$. We have seen no evidence for such high temperature oxygen 
burning; probably the best chance would be at the very onset of core collapse.

\begin{figure}[!hb]
\label{fig.8}
\psfig{file=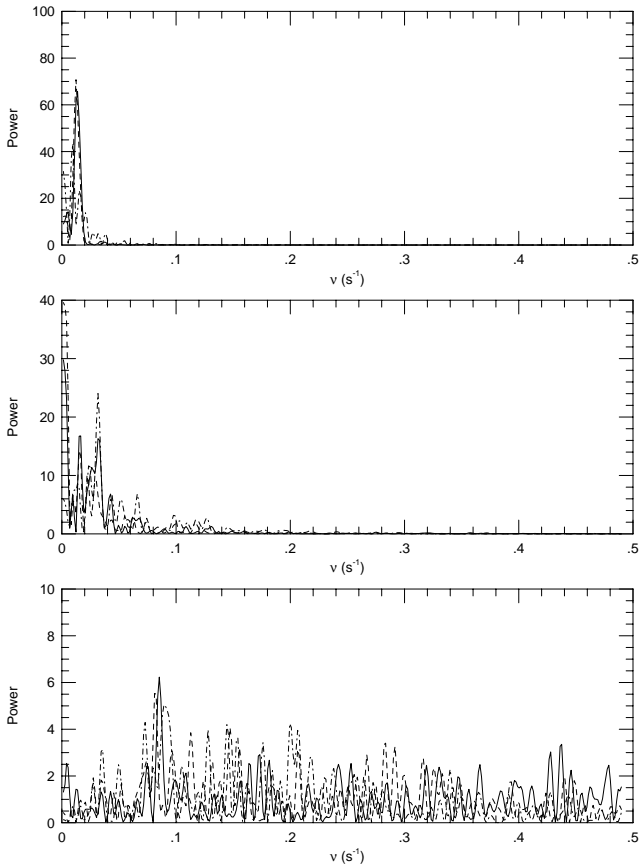,height=5.5in}
\caption{Power Spectrum over 200-400s, for top, middle and bottom 
of the convective region. At the top, most of the power is below
the Brunt-V\"assla frequency, indicating the importance of acoustic
and gravity modes. The spread in power over frequencies at the 
bottom (flame zone) is due to extreme time variability. }
\end{figure}

As with previous convection simulations (\cite{htm86}, \cite{htmz94}), we
conclude that gravity waves are responsible for the generation of the
perturbations. A fourier analysis of the power in vertical velocities
at different frequencies $\nu$, was made for the time interval from 200 
to 400s. Figure~8 shows the power as a function of frequency just 
above the convective zone (top panel), in the middle of the zone,
 and just below the convective zone (bottom panel);
compare with Figure~4 of \cite{htm86}.  The solid, dashed, and dot-dashed
 lines were taken at ${1 \over 4},{1 \over 2},{3 \over 4}$ of the angular
domain. 
In the lower panel, we see the signature of a complex series of modes. 
The local Brunt-Vaisala frequency is $N_{BV} = 1.55 \rm s^{-1}$, which
is much higher than most of the power. As did \cite{htm86}, we interpret
this as a time variation due to internal gravity waves, recalling that such
waves have frequencies less than the buoyancy frequency. 

However, just above the convective zone (top panel), we find that the power is
also at frequencies just below the local Brunt-Vaisala frequency,
 $N_{BV} \approx 0.02 \rm s^{-1}$. The level of power here is
larger, unlike \cite{htm86}, where the power above the convection zone was
smaller than the power just below the convection zone.
The difference seems to be in the nature of the driving and damping. In
our simulation, the driving energy flux from the oxygen burning flame 
zone varies by factors of two at the base of the Schwarzschild unstable region. 
In addition, the energy release rate in the $\rm^{12}C$-rich plume exceeds that
from the oxygen flame. 
It has already been inferred that g-mode waves of high radial wave-number can 
appear at the interfaces between convectively stable and unstable regions 
(\cite{p81}, \cite{htm86}), and in this case we seem to have preferential
driving of these modes.

While the appearance of g-mode wave interactions
in this simulation is consistent other studies (\cite{p81}, \cite{htm86}), 
the details are different.  Thermal 
diffusion is not present in this calculation, so that any damping of linear 
internal waves is the result of numerical dissipation (shocks).  Pressure waves 
generated 
within the convective zone should have a constant horizontal wavenumber as 
their source.  These internal waves are known to undergo internal reflection 
when N(r) = $\omega$, where N(r) is the Brunt-V\"aissala frequency 
at radius $r$ (\cite{p81}).
We note that the detailed coupling of acoustic and gravity waves 
to the convection also depends on boundary conditions (\cite{htmz94}). 

The greatest power in the middle of the convective zone occurs at zero frequency
(center panel), as found by \cite{htm86}. However, despite our significantly
better numerical resolution, our peaks are less well developed. We attribute this
to the fact that our convection is much further from statistical relaxation.
This is a physical feature of the epoch of oxygen shell burning, not a limitation
of the simulation, as our previous discussion shows.

\begin{figure}[!hb]
\label{fig.9}
\psfig{file=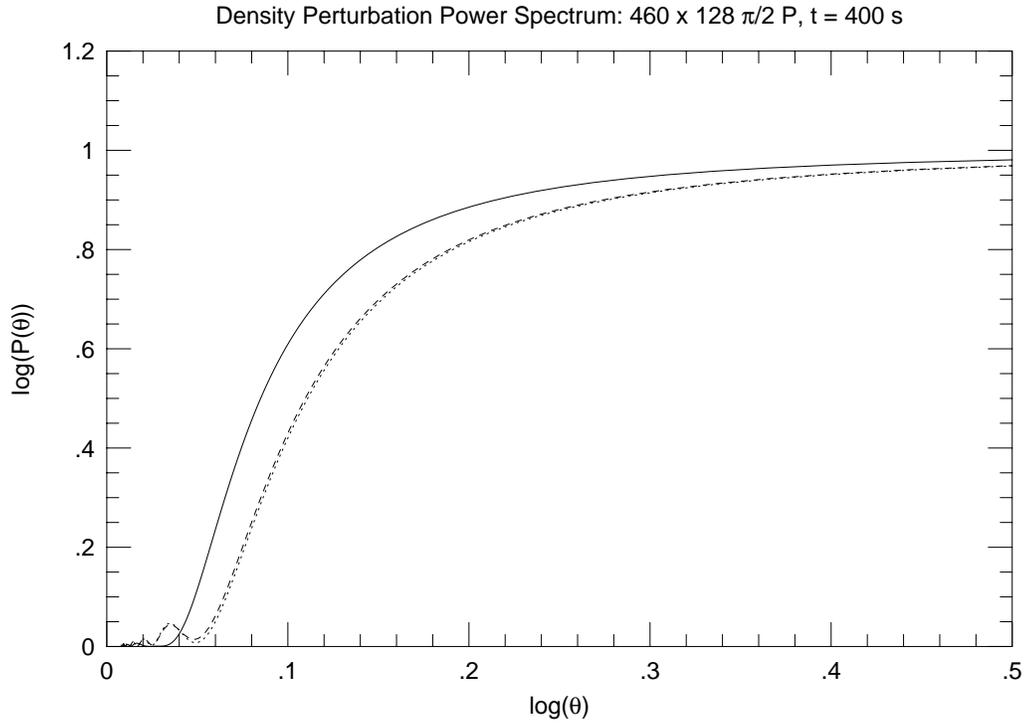,height=5.5in}
\caption{ Power Spectrum of density perturbations versus angular separations
$\Delta \theta$, at 400s.}
\end{figure}

Figure~9 shows the power spectrum of density perturbations versus angular 
separations $\Delta \theta$, at 400s. Three radii are indicated:
a solid line indicates a radius above the convective zone, a dot-dashed line
the convective zone, and a dashed line a radius below the convective zone.
There is a fall off at low angular separations,  $\Delta \theta \le 0.1$.
The zoning is much finer than that, and will allow clumping on a much
smaller scale ($1/n_\theta \approx 0.008$). This 
fall off could be taken as a measure
of the scale length of the perturbations to eventually be compared with
observations, and as such is an important quantity, but two dimensional
simulations such as these are biased. Because the vortices are pinned
on the grid---their angular momentum vectors stay perpendicular to the
plane of computation---cascading to smaller scales is inhibited. Thus we
expect clumping down {\em at least} to scales this size; this issue must
be addressed with fully three dimensional simulations.

\subsection{Additional Convective Mixing}

We mentioned that the asymmetry in the upflows and downflows has
direct consequences for convective mixing.  
In Figure~6, we showed a time
history of $^{12}$C abundance.  In the initial model, $^{12}$C 
exists at the 10$^{-7}$ abundance level in the oxygen convective zone,  and 
up to the 0.4 level above the zone.  As rising material overshoots into the 
high $^{12}$C abundance area, $^{12}$C-rich matter is entrained.
Eventually, cooling causes
fast moving downward plumes  to mix $^{12}$C deep into the convective 
zone.  These fingers terminate at the
point where $^{12}$C burns faster than it can be advected.  Nuclear energy
hot spots appear at these points, and by the end of the simulation the energy
output from these hotspots are a factor of 30 higher than the oxygen flame
zone itself.  Even $^4$He, which exists still farther away from the MLT 
convection boundary, is brought down into the convective shell, albeit at a 
low level.

\begin{figure}[!hb]
\label{fig.10}
\psfig{file=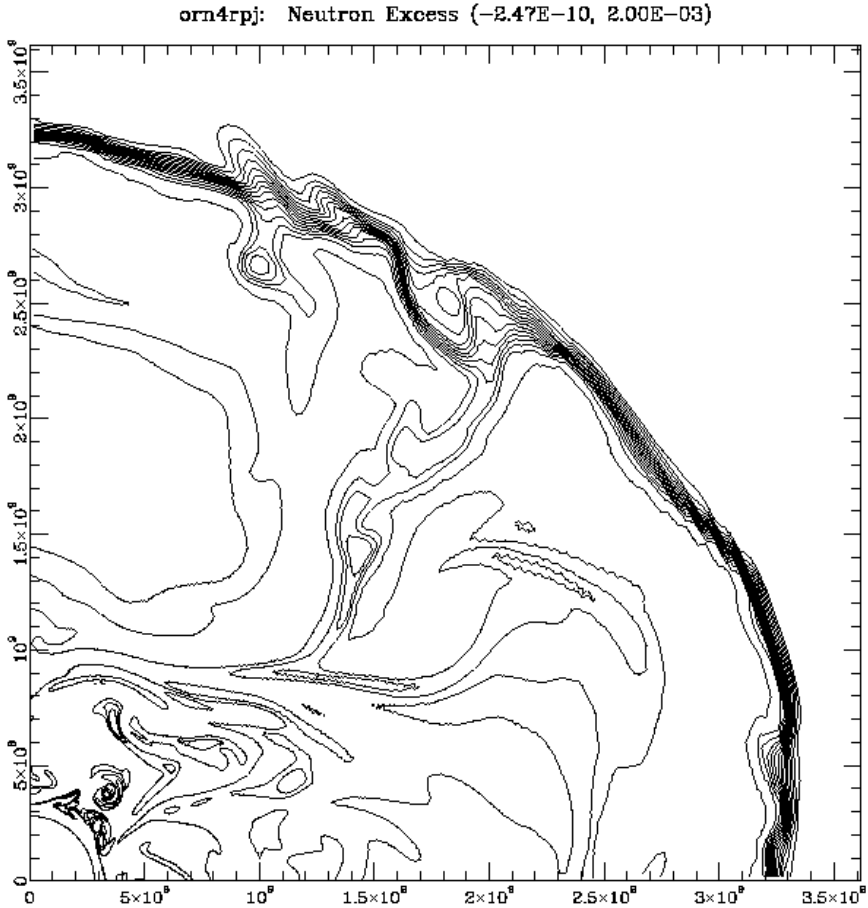,height=5.5in}
\caption{Contours of neutron excess $\eta$. 
Explosive nucleosynthesis yields are particularly sensitive to such
variations. Twenty contours are plotted, linearly spaced from
$\eta = 0 $ to $ \eta = 2 \times 10^{-3} $ .}
\end{figure}

An interesting phenomenon is the mixing of higher Y$_e$
into the convective shell. Figure~10 shows the perturbation structure of the 
neutron excess $\eta = 1 - 2 Y_e$ at the end of the calculation.  
The non-uniform distribution at both the upper and lower boundaries is
readily apparent.  Since the `mass cut'  between the proto-neutron star
and supernova ejecta is expected to occur within this convective zone, the
presence of a non-uniform distribution of Y$_e$ may alter core collapse
(\cite{bl85}, \cite{bf92}).  A larger effect could be from the
coupling of the electron capture with hydrodynamics, which would change
the entropy and the size of the core (\cite{auf93}, \cite{a96}).
Furthermore, once the delineation
between neutron star and ejecta is made, explosive nucleosynthesis in the
ejecta will reflect the distribution of neutron excess $\eta$.
The differences of $\Delta \eta \sim 0.002$ that we find here 
will radically alter the details of the quasi-statistical equilibrium (QSE) 
nuclear burning process (\cite{tcg66}, \cite{mf72}, \cite{wac73}) so that 
a non-uniform distribution of isotopes is produced. This may be a welcome 
result, as one dimensional models of explosive burning in massive stars have
usually produced too many neutron-rich isotopes (\cite{hcaw74}, \cite{hwe85}),
unless the mass cut was carefully chosen (e.g., \cite{tn90}).

Most significantly, this extensive mixing by plumes may indicate that
our concept of convection is too narrow to properly describe the
neutrino cooled stages of massive star evolution. 
If this stage is so drastically altered, why not earlier stages as well?

\subsection{Stellar Structure and Evolution}

We have found that the results of our two-dimensional computations are 
inconsistent with 1D hydrostatic MLT results in many ways.  
As yet the evolution has proceeded about 10~\% of the time from
oxygen shell ignition to core collapse  (for the 1D evolution of this
initial model). What can we infer about 2D effects on evolution?

In 1D models, under MLT conventions,
the O-burning region is wholly contained within the convective zone.  
This determines the evolution: the convective region burns until it
changes its fuel to ashes, and there is little ingestion of new fuel
or loss of ashes.
Thus, at the exhaustion of fuel in the convective region, this layer
of ashes is added to the underlying core. Contraction follows, and
a new shell generally ignites just above the old, 
now defunct convective shell, and the process is repeated. 
This behavior has long been known (\cite{rs67,a69a}); see
\cite{a96}, \S10.1 for details.
\cite{a72} noted its possibly non-physical nature, contrasted it
to the case of a radiatively cooled shell, and also
considered an alternative in which the convective region both lost
ashes and ingested new fuel, so that it burned out in mass coordinate
as a ``wave.''  However, we are aware of no
published evolutionary sequences of neutrino cooled shells 
except those calculated with the implicit assumption that
the first picture is correct. This distinction involves the
nature of the {\em boundaries} of the convective region.

In these 2D simulations,
within the first 25 seconds of evolution time, the flame zone `flashed' and
changed its thermodynamic state from convective instability to 
stability, thereby removing itself from the convective zone, as 
MLT would define it.  The resulting sound wave does not develop into a shock as
it moves down the density gradient.  No structure changes occur by the 
passage of this wave. 

\begin{figure}[!hb]
\label{fig.11}
\psfig{file=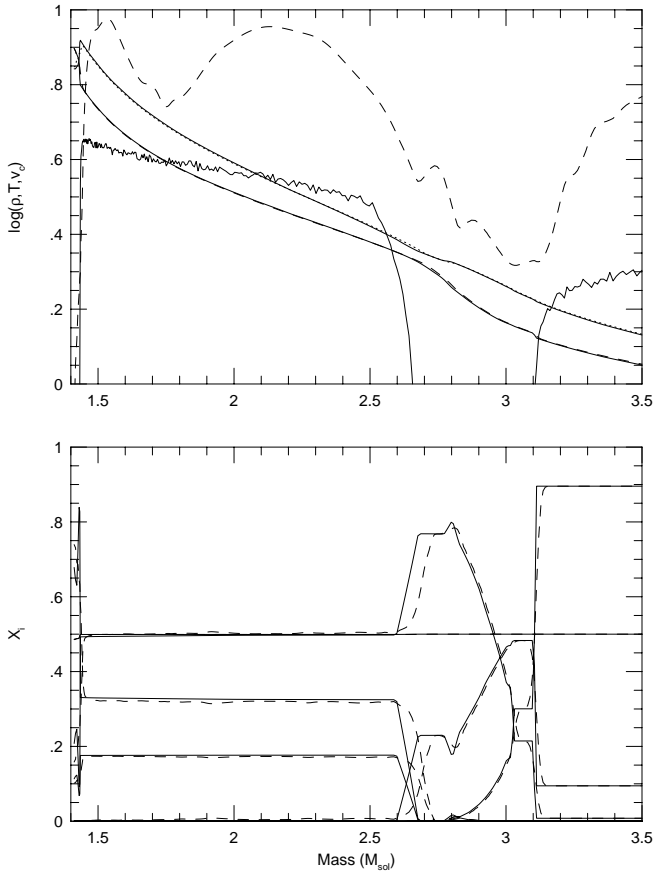,height=5.5in}
\caption{Top: Azimuthally averaged $\epsilon$, $\rho$, and T, at 0 and 400s. 
Bottom: same for composition. While the gross structure is preserved, the
flame zone changes. The velocity field is qualitatively different from 1D, 
allowing mixing across stable regions.}
\end{figure}

The upper panel of Figure~11 shows the azimuthally averaged values of
energy production, temperature, density, and convective speed,
as a function of mass coordinate, for the initial
and final state in this simulation (that is, at 0 and 400s).  
The variables of the initial model are represented by solid lines; those
of the final model by dashed and by dotted lines. The most obvious
difference is in the convective speed. In the initial (MLT) model, the
velocities are significant only in the convective oxygen burning shell
and the convective helium burning shell. They are separated by a stable
region containing C, Ne, Mg, and mostly O. In the final model, the
convective speed is larger, and does not go to zero in this intermediate
region. While in MLT the convective speed is a slowly varying quantity,
in 2D simulations it fluctuates strongly. 

This additional mixing has affected the composition structure, which
is shown in the lower panel of Figure~11. This is especially noticeable
in the range $2.5 < M_r/M_\odot < 3.2$. It is interesting that the
He burning shell, which is broadened by radiative diffusion, widens
more due to convection. 

A second important difference is in the thin flame zone, which is
strongly compressed in the mass coordinate, and therefore less obvious
in Figure~11. Consider the temperature and density structure. 
Just above the inner boundary, the temperature has a small
hole and the density a corresponding peak.
At first sight, the run of both variables is almost idential between
the initial and final models, but in fact the {\em boundary} has changed. 
The steep abundance gradient below the oxygen flame zone
is becoming shallower, and the oxygen flame is no longer in the
most convectively active region. Similarly, the
outer edge of the temperature ``hole'' is smoothed out. 

\begin{figure}[!hb]
\label{fig.12}
\psfig{file=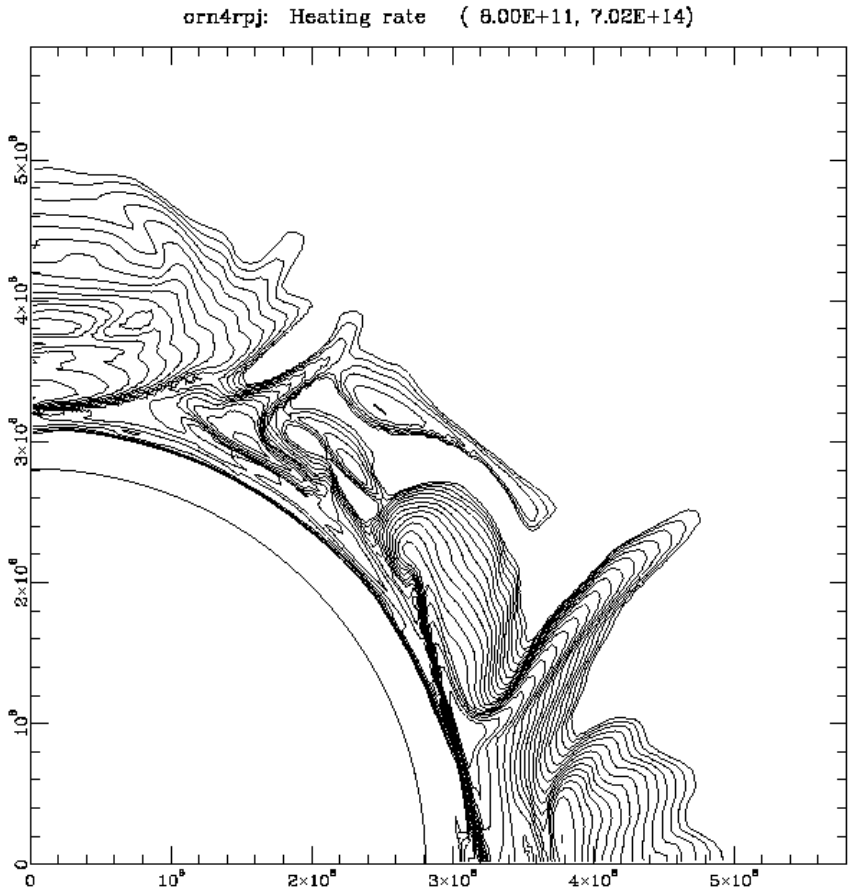,height=5.5in}
\caption{Closeup of Flame Region: Contours of the total rate of
nuclear heating.}
\end{figure}

Figure~12 shows a closeup of the heating rate in the flame region.
There is one velocity vector for each 25 zones. The 20 density 
contours are logarithmically spaced between $\epsilon = 8.0 \times 10^{11}$
and  $7.0 \times 10^{14}\rm erg\ g^{-1}\ s^{-1}$. The oxygen burning
shell is burning at  
$\epsilon \le 1.3 \times 10^{13}\rm erg\ g^{-1}\ s^{-1}$,
and so is barely discernable as an inner ledge. The oxygen burning is not
uniform, having several peaks.

Most of the energy release comes from carbon and neon burning, as blobs 
having $X(\rm ^{12}C)$ of a few percent burn vigorously. These conditions
($T\approx 2.3 \times 10^9 \rm K$ and $\rho \approx 8 \times 10^5 \rm g/cc$),
occurring on a hydrodynamic time scale, are typical of ``explosive
nucleosynthesis'' of carbon and neon. 

Notice that
the ``hot spots'' correspond to downward moving matter. The width of
the whole flame zone is about 5 times wider that of just $X(\rm ^{16}O)$,
and much more uneven. The velocities in the fastest burning zones
(both oxygen and carbon) are relatively small, so the ashes are not
immediately swept out into the most active convection.

This situation is drastically different from the conventional picture
suggested by 1D models. It causes worry that the 2D evolution would have
diverged from the 1D at a time even earlier than that for our initial
model. We are currently attempting to push this sequence to later times
(core collapse).  

\subsection{Dimensionality}

Two dimensional simulations such as these clearly are more realistic
than one dimensional simulations with ad hoc prescriptions for convection
and mixing. Hydrodynamic flow is simulated self-consistently, and
heterogeneity in space and time may be examined. Nevertheless, the
assumed ``rotational'' symmetry (not presumed due to rotation but
simply enforced by limitations in computational resources), is simply
not valid. Simulation of a 3D sector, for example by extending the
460x128 by 64 zones in $\phi$, is feasible on a massively parallel
supercomputer. We are currently continuing this calculation with a PVM 
version of PROMETHEUS, and will to extend the simulations 
to three dimensions by the time of publication.

It is unclear at this time what physical state will be described by a three 
dimensional simulation. It is well known that the energy 
cascade in two dimensional turbulence (lower to higher scales) is opposite
that in three dimensions (higher to lower scales).  This effect can be seen
quite easily in our and other PPM-based simulations (\cite{pw94}), where 
small eddies disappear in favor of larger scale eddies.  A recent comparison
of two- and three-dimensional simulations of convection in a proto-neutron star
(\cite{mj97}) also show an increase of small scale features in the 
three-dimensional simulation relative to the two-dimensional simulations, with
an overall lower kinetic energy in the three-dimensional case.  Thus, the 
overall convective iand composition fields are expected to be more homogeneous
in a three-dimensional case.  

However, the effects of a third dimension on overshoot are not at all clear.
This is critical for the oxygen shell evolution in light of what we mentioned
previously concerning carbon brought into the oxygen shell by overshoot and
the inhomogeneous energy release.  On the one hand, we know that overshoot
is the result of individual plumes creating gravity waves at convective
stability$/$instability boundaries.  Such plumes are known to be more energetic
in three dimensions than two due to a reduced drag effect (two dimensional
`plumes' are tori in realty) if starting from the same initial conditions.
We know, though, that the initial underdensity which characterizes a plume is a 
function of the kinetic environment, which, from the supernova simulation 
mentioned above and turbulence studies, should be reduced from its two 
dimensonal analog.  Just which effect will predominate requires a three 
dimensional simulation.  More insight can be obtained from laser experiments 
for related hydrodynamic problems, where Rayleigh-Taylor instabilities in 2D 
and 3D are examined experimentally within the context of inertial confinement
fusion (\cite{br95}). Testing of PROMETHEUS against ICF codes and
NOVA laser experiments is beginning (\cite{jk96}). 

\section{CONCLUSIONS}

Direct calculation of the hydrodynamic behavior of oxygen burning shells
indicates that the process is more interesting than previously supposed.
Previous 2D work (\cite{a94,ba94}) is confirmed and extended. 

Composition is not homogeneous in the convective region. The system has
strong fluctuations in both space and time. Mach numbers in the flow
are typically 0.1 to 0.2; pressure fluctuations and acoustic fluxes are
not negligible. Any attempt to model systems such as these with implicit
or anelastic methods will miss the significant contribution of energy
transport by sound waves. This convective flow will not reach
a statistical steady state prior to core collapse. While the convective
motions are not sufficient to alter the stellar structure significantly
through their Reynolds stresses, they do allow mixing of composition from the
convectively stable region into the unstable ones. The evolutionary
implications of this have not yet been explored.

In light of these results, we view with skepticism the details of one
dimensional stellar evolution calculations for this stage
(e.g., \cite{ww95,tn90}). 

Prior to the core collapse, perturbations in density develop in the
oxygen shell which are sufficiently large to ``seed'' hydrodynamic
instabilities which will mix the ``onion skin'' composition of the
presupernova. This occurs in precisely the region in which $\rm ^{56}Ni$
is explosively produced by oxygen burning behind the explosion shock.

Although the active heating and cooling by nuclear burning and neutrino
emission is novel, we find good agreement with previous multi-dimensional
simulations of convection. Our results seem insensitive to numerical
details.

\acknowledgments

Enlightening discussions with Prof. Phillip Pinto and Willy Benz 
are gratefully acknowledged.  Information from Nick Brummell about
work in progress is sincerely appreciated.  This work was supported 
in part by NASA grant NAGW-2450, and by NSF grant ASTRO-9015976.

\clearpage
\appendix
\section{NUMERICAL EXPERIMENTS}

\begin{figure}[!hb]
\label{fig.13}
\psfig{file=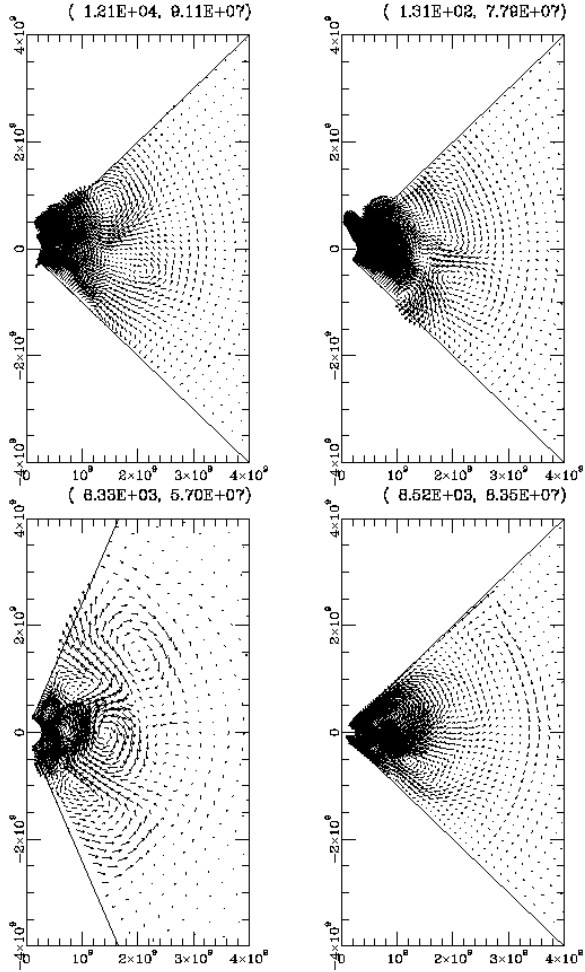,height=5.5in}
\caption{Velocity vectors for Models D (upper left), B (upper right),
E (lower left), and C (lower right). See Table~1.}
\end{figure}

Figure~13 shows the velocity vectors for several numerical experiments:
Models D, B, E, and C from Table~1. The two top panels show a change in 
resolution by a factor of 2 in each linear dimension. The higher 
resolution case is to the left. In the high resolution model, velocity vectors
are the average vector over 4 zones, while the vectors in the lower resolution 
models of 90 and 120 degree extents are 2 and 3 zone averages, respectively.
The flow patterns are qualitatively the 
same, and show some quantitative similarity as well, which leads us to 
believe that this resolution is the minimum needed for a `converged'
solution.  The lower left panel shows 
the effect of widening the wedge from 90 to 120 degrees; we now have three 
vortices in the midlayer instead of two and a half. A larger difference may 
be seen in the lower right panel, which used reflecting boundary conditions.
The vortices are all contained within 90 degrees. Despite this, the
qualitative features are still similar.

\begin{figure}[!hb]
\label{fig.14}
\psfig{file=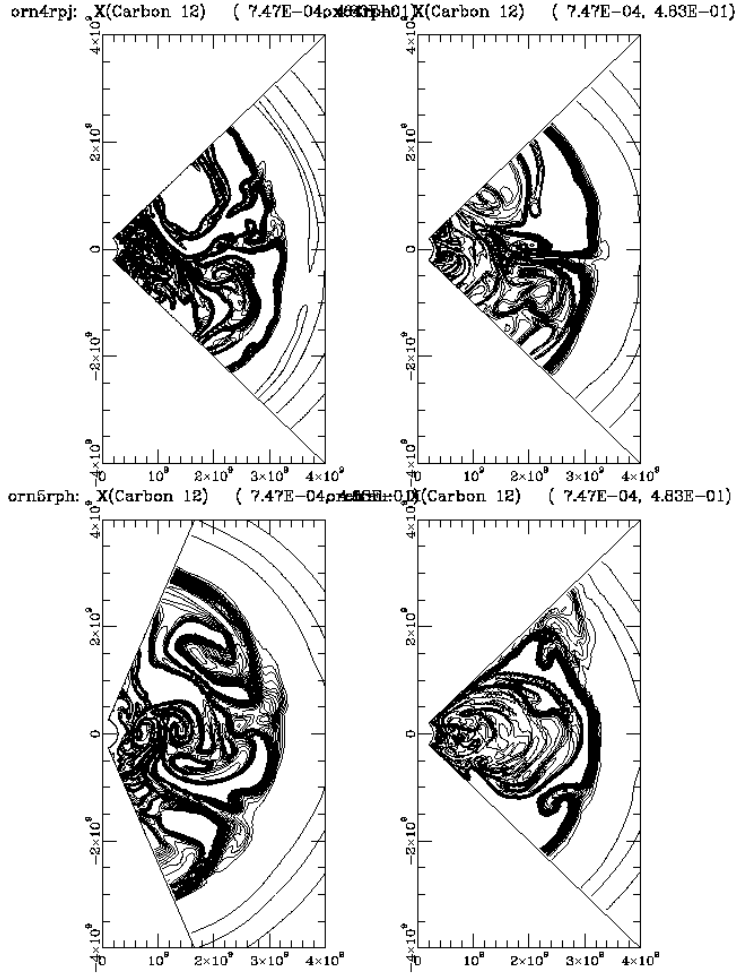,height=5.5in}
\caption{Abundance contours for $^{12}$C for models D, B, E, and C, as
in Figure~13. See  Table~1.}
\end{figure}

Figure~14 shows the corresponding panels for contours of $^{12}$C abundance.
In all cases we have vigorous burning of an entrained plume of $^{12}$C-rich
matter.


%
%
%
%

\clearpage

\clearpage

%
%

%

\clearpage

\end{document}